# Fluctuation analysis in nonstationary conditions: single Ca channel current in cortical pyramidal neurons

C. Scheppach, H. P. C. Robinson

Running title: fluctuation analysis: Ca channel current

## Abstract

Fluctuation analysis is a method which allows measurement of the single channel current of ion channels even when it is too small to be resolved directly with the patch clamp technique. This is the case for voltage-gated calcium channels (VGCCs). They are present in all mammalian central neurons, controlling presynaptic release of transmitter, postsynaptic signaling and synaptic integration. The amplitudes of their single channel currents in a physiological concentration of extracellular calcium, however, are small and not well determined. But measurement of this quantity is essential for estimating numbers of functional VGCCs in the membrane and the size of channel-associated calcium signaling domains, and for understanding the stochastic nature of calcium signaling. Here, we recorded the voltage-gated calcium channel current in nucleated patches from layer 5 pyramidal neurons in rat neocortex, in physiological external calcium (1-2 mM). The ensemble-averaging of current responses required for conventional fluctuation analysis proved impractical because of the rapid rundown of calcium channel currents. We therefore developed a more robust method, using mean current fitting of individual current responses and band-pass filtering. Furthermore, voltage ramp stimulation proved useful. We validated the accuracy of the method by analyzing simulated data. At an external calcium concentration of 1 mM, and a membrane potential of −20 mV, we found that the average single channel current amplitude was about 0.04 pA, increasing to 0.065 pA at 2 mM external calcium, and 0.12 pA at 5 mM. The relaxation time constant of the fluctuations was in the range 0.2–0.8 ms. The results are relevant to understanding the stochastic properties of dendritic $Ca^{2+}$ spikes in neocortical layer 5 pyramidal neurons. With the reported method, single channel current amplitude of native voltage-gated calcium channels can be resolved accurately despite conditions of unstable rundown.

## Introduction

When an ion channel opens, a characteristic fixed current flows through it, which can be directly measured with the patch clamp method (1). Even when the single-channel current is too small to be resolved with this technique, it can still be inferred by fluctuation analysis, where the current through many channels is measured and the fluctuations arising from the opening and closing of individual channels are analyzed (2–5).

Voltage-gated calcium channels (VGCC) (6–8) are central to many physiological processes, and are present in most neuronal membranes, where they have important roles in signaling. The apical dendrites of neocortical layer 5 (L5) pyramidal neurons (9, 10) exhibit action potentials carried by $Ca^{2+}$ (11). These dendritic $Ca^{2+}$ spikes provide an amplification mechanism for distal synaptic inputs, triggering bursts of sodium action potentials at the soma and hence constitute a fundamental mechanism of synaptic integration in these neurons (12, 13). Similar dendritic calcium spikes have been found in hippocampal CA1 pyramidal cells (14), and Purkinje neurons (15). At presynaptic terminals, calcium influx through voltage-gated calcium channels triggers the vesicular release of neurotransmitter (16).

The amplitude of currents through native single $Ca^{2+}$ channels at physiological extracellular $Ca^{2+}$ concentrations (1-2 mM) (17) is a key parameter. Knowing it is essential to be able to relate the density of functional channels to the size of whole-cell calcium currents, to characterize the stochasticity of calcium signaling due to random opening and closing of single ion channels (18–20), and to understand the formation of calcium nanodomains, highly localized and concentrated plumes of calcium on the cytoplasmic side of open calcium channels, allowing specific signaling (21), which depends on the rate of $Ca^{2+}$ entry through individual channels.

However, physiological single $Ca_V$ channel current has not been extensively investigated, with only a few estimates in the literature (22, 23, 21, 24). In chick dorsal root ganglion neurons, values between 0.2 and 0.33 pA were estimated for different VGCC subtypes (21), at a membrane potential of −65 mV and an extracellular calcium concentration of 2 mM, by direct single-channel recording under sophisticated low-noise conditions, and by extrapolating from measurements at more elevated calcium concentrations. Such a small amplitude makes it extremely challenging to resolve and estimate accurately. Single calcium channels are therefore studied almost exclusively in high extracellular barium solutions (e.g. (25)), since barium permeates most VGCC subtypes at higher rates than does calcium, giving much larger, detectable channel current amplitudes close to 1 pA in size.

To better understand the stochastic properties of $Ca^{2+}$ spikes in neocortical L5 pyramidal neurons, we wanted to measure the single channel current of the native $Ca^{2+}$ channels in these cells, in physiological $[Ca^{2+}]_o$, for which fluctuation analysis appeared a suitable approach. However, because of rapid run-down of the channel current (a common problem in the study of $Ca^{2+}$ channels, see (26, 27)), fluctuation analysis cannot be applied in its usual form, relying on stable conditions to obtain mean current and current variance by ensemble-averaging across several recording sweeps. We therefore developed a modified fluctuation analysis protocol which extracts information from individual sweeps, based on mean current fitting and band-pass filtering. Furthermore, we used voltage ramp stimuli rather than the more customary voltage steps. The method was validated by modeling. We obtain a mean single $Ca^{2+}$ channel current of 0.065 pA at −20 mV and 2 mM extracellular calcium, which is about one order of magnitude below the resolution limit of standard patch-clamp single-channel recordings.

## Methods

### Tissue preparation

Animal procedures adhered to U.K. Home Office legislation and University of Cambridge guidelines. Acute brain slices were obtained from Wistar rats aged 6-14 days. Animals were killed by dislocation of the neck, and parasagittal brain slices were prepared as detailed in (28). The brain was cut along the midline and one hemisphere was glued by the medial surface to a tilted block (12°) on top of the horizontal slicer stage, slightly raising the ventral part of the brain. 300 μm thick slices were obtained with a vibratome (VT1200S by Leica, Milton Keynes, U.K.), with the dorsal cortical surface facing the horizontal blade. Slices were incubated at 34°C for 30 min and then kept at room temperature.

### Solutions, drugs, chemicals

Extracellular solution (used for slicing and perfusion of slices) had the following composition: 125 mM NaCl, 2.5 mM KCl, 2 mM $CaCl_2$, 1 mM $MgCl_2$, 1.25 mM $NaH_2PO_4$, 25 mM $NaHCO_3$, 25 mM glucose, bubbled with carbogen gas (95% $O_2$, 5% $CO_2$), pH 7.4 (c.f. (29) p. 200, (30, 28)). Intracellular Cs-based pipette solution had the following composition: 90 mM Cs-methanesulfonate, 30 mM CsCl, 10 mM BAPTA-$Na_4$ ($Ca^{2+}$ chelator), 11 mM HEPES buffer, to which was added 4 mM ATP (ATP Mg salt obtained from Sigma, Gillingham, U.K.; free Mg: about 1 mM), 0.3 mM GTP (GTP Na salt hydrate obtained from Sigma; 1.2 mM $Na^+$), and 10 mM creatine phosphate (CrP) $Na_2$ (Na CrP dibasic tetrahydrate obtained from Sigma), to mitigate run-down of $Ca^{2+}$ channels (26), and pH was adjusted to 7.3 with NaOH and HCl (about 4.5 mM NaOH for HEPES, about 8 mM NaOH for ATP). The liquid junction potential of 12.2 mV was corrected for. HEPES-buffered extracellular solution consisted of: 145 mM NaCl, 2.5 mM KCl, 2 mM $CaCl_2$, 1 mM $MgCl_2$, 10 mM HEPES, pH adjusted to 7.4 with about 4.2 mM NaOH. Tetrodotoxin (TTX, Na channel blocker) was obtained from Alomone Labs (Jerusalem, Israel) and Tocris (Bristol, U.K.).

**Recording**

Slices were viewed with an Olympus (Southend-on-Sea, U.K.) BX50WI fixed stage upright microscope and a x60 objective, or a x10 objective for larger-scale overview, with infrared (IR) or visible light differential interference contrast (DIC) optics and a Lumenera (Ottawa, Canada) Infinity 3M CCD camera. To assist orientation in the slice, the position in the slicing plane was monitored with an optical position encoder (Renishaw, Wotton-under-Edge, U.K.), displaying x and y coordinates at micron resolution. Pipette pressure was monitored by a pressure meter. Voltage clamp data were collected with a MultiClamp 700B (Axon Instruments, Molecular Devices, Wokingham, U.K.) amplifier, with a feedback resistor of 50 G$\Omega$ and a 10 kHz 4-pole Bessel filter, with waveform generation and sampling at 16-bit resolution at a frequency of 50 kHz, using a National Instruments (Newbury, U.K.) analog interface and custom software written in C++ and Matlab (Mathworks, Cambridge, U.K.). During the experiment, perfusion with extracellular solution was maintained in the recording chamber with a perfusion line and a suction line. All experiments were performed at room temperature (20-23°C).

Pyramidal neurons in cortical layer V were visually identified. In order to minimize variability, cells from the same cortical location were consistently targeted. Slices were always obtained from the right hemisphere. In consecutive parasagittal slices, the orientation of pyramidal cell dendrites changes from slice to slice, and slices in which the dendrites run roughly parallel to the slice surface were selected, which came from about half-way between the midline and the lateral brain surface. The central part of the cortex in the rostral-caudal direction was targeted, corresponding to the sensory hindlimb region (31).

Patch pipettes were pulled from borosilicate glass capillaries (Harvard Apparatus, Cambridge, U.K.; 1.5 mm outer diameter, 0.86 mm inner diameter, filamented glass) and fire-polished. Tip resistances (with the Cs-based pipette solution) were between 5.5 and 6.5 M$\Omega$. Nucleated patch recordings were performed as described in (32). A whole-cell recording was obtained, and a negative pressure of about −200 mbar was applied to the pipette, aspirating the cell nucleus to the pipette tip. Maintaining the negative pressure, the pipette was slowly and carefully retracted, such that the nucleus was pulled out of the cell and acted as a place holder around which the cell membrane resealed. The nucleated patch was pulled out of the slice and then raised well above it. Finally, the negative pipette pressure was reduced to about −50 mbar. The process from initial contact of the membrane with the pipette tip to the first recorded sweep took about 4 to 7 minutes.

$Ca^{2+}$ channel currents were recorded during depolarizing ramps, from a holding potential of −70 mV rising to 0 mV (slope 0.6 mV/ms, stimulus repeated every 3 s), and also during depolarizing voltage steps (holding potential −70 mV, 100 ms long steps to test potentials cycling through −60, −40, −30, −20, −10 and 0 mV, stimulus repeated every 3 s).

**Data analysis, modeling and statistics**

Analysis was carried out using custom code in MATLAB and R. Leak subtraction for step responses (Fig. 1A) was carried out by subtraction of the smoothed, scaled response to a small step (holding potential +10 mV) below the activation range of the calcium current. All current traces were digitally filtered with a low-pass Gaussian filter ((29) p. 485) at a −3dB corner frequency $f_c$ = 1 kHz. To isolate current fluctuations, voltage-ramp-evoked currents were fitted with a piecewise polynomial, consisting of a flat segment and two linear segments linked by cubic splines (see Supporting Material), giving a smooth, continuous-derivative heuristic function whose eight parameters were fitted by minimizing squared deviation from the data, using a Nelder-Mead simplex algorithm (Matlab function fminsearch), which was taken as the estimate of the mean current $\mu_I$. Following subtraction of this function, the fluctuating current responses were passed through a high-pass filter, by subtracting from the signal its low-pass-filtered version (Gaussian filter with $f_c$ = 50 Hz, resulting in a high-pass filter with corner frequency 94 Hz), to reject extraneous low-frequency noise and error introduced at low frequencies by the mean-fitting procedure. The variance of the resulting band-pass filtered fluctuations ($\sigma_I^2$) was computed over brief (5 ms) time bins, and the single channel current $i$ estimated from the relationship $i \approx \sigma_I^2/\mu_I$ (see Results).
Weighted straight-line fits with fit parameter estimates, errors and correlations were obtained with the statistics package R.

## Results

### Calcium channel currents in nucleated patches

To perform fluctuation analysis effectively, one needs a large membrane containing as many channels as possible, but which can be voltage-clamped reliably. Normal whole-cell voltage-clamp recording is impractical in pyramidal neurons because the large apical dendrite leads to space clamp problems. Therefore, we used the nucleated patch technique, which combines a large available membrane area with excellent space clamp properties.

Voltage-gated $Ca^{2+}$ channel currents were recorded in nucleated patches from neocortical L5 pyramidal neurons, using a cesium-based intracellular pipette solution to block potassium channel currents (26), 1 μM TTX in the Ringer solution to block sodium channel currents, and 2 mM extracellular calcium. Figure 1A shows an example family of leak-subtracted currents in response to step depolarizations. The current is activated around −30 mV and has a more pronounced inactivating component as the membrane is depolarized further. The small amplitude of these currents at physiological calcium concentration means that the signal-to-noise ratio is also small. Responses to ramp depolarization (Fig. 1B) also clearly show the onset and voltage-dependence of the current. Application of 200 μM cadmium completely abolished the current (Fig. 1C, $n$ = 11 cells), allowing its unambiguous identification as a voltage-gated calcium channel current. As expected for calcium currents (26, 27), despite inclusion of ATP and GTP in the pipette, there was often a relatively fast rundown over the first 5-10 minutes following establishment of the recording (Fig. 1B).

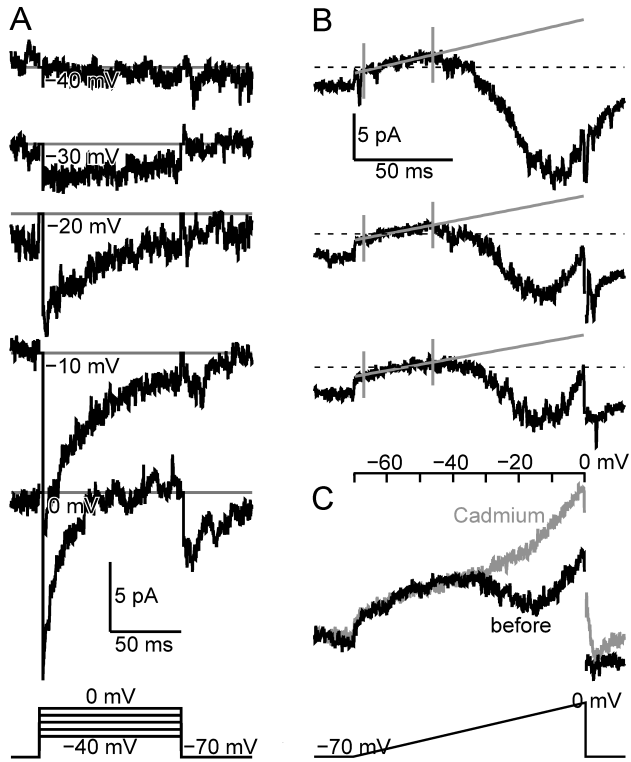

FIGURE 1  Voltage-gated $Ca^{2+}$ channel currents in nucleated patches from L5 pyramidal neurons. A) Leak-subtracted inward voltage-gated $Ca^{2+}$ channel current during a family of voltage steps from −70 mV to −40, −30, −20, −10, 0 mV. B) Three current responses during voltage ramps from a holding potential of −70 mV to 0 mV. During the initial part of the ramp, there is a linear leak current (indicated by grey lines; grey vertical lines indicate the range of the straight-line fit, from 5 ms to 40 ms after ramp onset). From about −40 mV, an inward current develops. (Ramp stimulus was repeated every 3 s. The three responses shown are each 21 s apart. Dashed horizontal line: zero current.) C) The inward current is eliminated 18 s after addition of extracellular $Cd^{2+}$ (200 μM).

**Fluctuation analysis in slowly nonstationary conditions**

Single calcium channel openings at physiological calcium concentrations are effectively impossible to resolve directly, because the combination of their small amplitude and short open duration means that their signal falls below the noise level of the patch-clamp technique. With the fluctuation analysis method, it is possible to infer the single channel current nevertheless, by measuring the current through many channels and quantifying its fluctuations arising from the opening and closing of individual channels (2–5). Assuming $N$ identical, independent channels with open probability $p$ leads to the following relationship of total current variance $\sigma_I^2$ to the ensemble mean level of current $\mu_I$:

$$\sigma_I^2 = i\mu_I - \frac{\mu_I^2}{N}, \qquad (1)$$

where $i$ is the single channel current (review: (29) p. 81, (6) p. 384). If $p$ is small, then

$$\sigma_I^2 \approx i\mu_I. \qquad (2)$$

Fluctuation analysis of voltage-gated channels is usually done by measuring channel currents during steps to a constant voltage (4, 5). Current mean and variance are obtained by ensemble averaging a number of consecutive sweeps. Computing a suitable ensemble average relies on complete stability of conditions and number of available channels over a considerable number (at least ≈20) of consecutive sweeps. However, this is not possible in the present case of native voltage-gated calcium channel currents. The channel current runs down visibly within one minute (Fig. 1B). In addition, the resistance and reversal potential of the linear leak current are not perfectly constant from sweep to sweep (Fig. 1B). While these instabilities constitute only very low-frequency variability and occur despite a good noise level at medium frequencies, they thwart the customary approach of obtaining current mean and variance from multiple successive ensemble trials.

We therefore aimed to extract information from individual recording sweeps. For that, we obtained the mean current not by averaging several successive sweeps, but by fitting a smooth curve to the current trace. The approach is similar to that used for quantally-varying nonstationary synaptic currents (33, 34).

Secondly, we found it advantageous to use not voltage steps, but voltage ramp stimuli, which allows easy leak subtraction by fitting a straight line to the initial part of the ramp response, produces a smooth mean channel current time course that can be well fitted, and naturally samples the full voltage range where the channels are active.

**Fluctuation analysis of voltage ramp responses**

The leak current was obtained from a straight line fit to the initial part of the ramp response from 5 ms to 40 ms after the start of the ramp (Fig. 1B), corresponding to a potential range of −67 to −46 mV, below the activation range of the calcium current. After subtraction of the linear leak current, to obtain the mean current, each ramp response was fitted by least-squares (see Methods) with a piecewise polynomial function (Fig 2A, see Methods). This heuristic function gave very good fits to current traces.

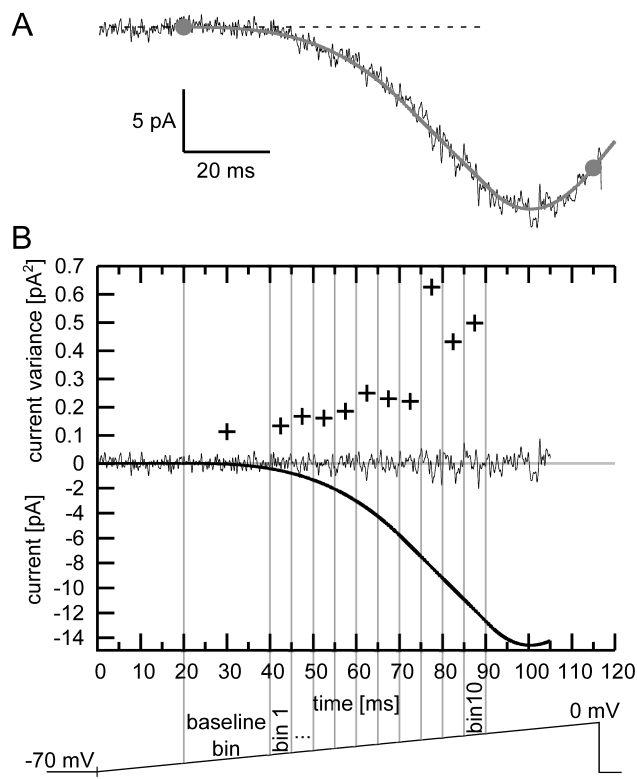

FIGURE 2  Isolation of current fluctuations. A) A heuristic piecewise polynomial function with continuous derivative (see methods) was fitted by least squares to individual leak-subtracted ramp responses (black trace). There were no systematic deviations of fits (grey curve) from individual current traces. Grey dots indicate the range of the fitted data. B) Current variance (black crosses) was computed over 5 ms time bins through the current

fluctuations (central black trace, obtained by subtracting from the current trace (panel A) the fitted mean current, and high-pass filtering). Black curve in the lower part indicates the mean current fit.

We then applied high-pass filtering to the mean-subtracted responses. This step increased the robustness of the measurement, by confining the signal to a bandwidth (94 Hz - 1 kHz) containing a substantial fraction of the channel noise power, while rejecting potentially large fluctuations of baseline at low frequencies. Secondly, it dealt with the problem that fitting the mean current to each individual ramp response leads to an underestimation of the current variance (see also Discussion). Since we subtract from the data not the underlying true mean current, but the fitted mean current, the obtained fluctuation trace is contaminated by their difference. This difference contains only low frequencies and is hence strongly attenuated by the high-pass filter. Fig. 2B shows an example of how the variance of the filtered fluctuations $I_{\text{residual}}(t)$, computed as the mean of $I_{\text{residual}}(t)^2$ over a time bin, evolves during an individual current response to a voltage ramp.

The drawback of high-pass filtering the fluctuations is that it also removes some of the fluctuation power due to channel gating and therefore the variance must be corrected for this. To that end, we investigated the spectral composition of the fluctuations, so that the effect of filtering could be quantified. We found that the autocorrelation function of the fluctuations was well-fitted by a single exponential relaxation (Lorentzian frequency component) passed through the same filtering (see Supplemental Material, Fig. S2A) with noise time constant $\tau$ in the range 0.2-0.8 ms (detailed data: Supplemental Material, Fig. S2B). The combined effect of high- and low-pass filtering cuts a substantial amount of the power of channel gating fluctuations (see Supplemental Material, Fig. S2C), resulting in a correction factor $\gamma$ by which the current variance is reduced, i.e. $\gamma = \sigma_{\text{filtered}}^2/\sigma^2$, where $\sigma_{\text{filtered}}^2$ is the experimentally measured, filtered current variance and $\sigma^2$ is the full variance of the unfiltered channel current fluctuations. $\gamma$ is a function of $\tau$ and can be calculated by integrating the filtered Lorentzian power spectral density over frequency, and dividing by the total variance (see Supplemental Material, Fig. S2D). $\gamma$ is actually rather insensitive to $\tau$, remaining between 0.5 and 0.574 over the measured range of $\tau$. Thus the measured variance is converted to true total variance by dividing by $\gamma$.

Using ramps means that the voltage is not constant, but the single channel current $i$ depends on voltage. Therefore, fluctuation analysis has to be performed in each of the ten time bins separately. In the end, the results can be re-combined in a single-channel current-voltage plot. Applying this to actual data (Fig. 4) using one patch as an example, the baseline noise variance $\sigma_{\text{baseline}}^2$ was measured in many repeated trials over the 20-40 ms time bin (Fig. 4A), as well as the variance $\sigma_{\text{exptl}}^2$ in bins 1 to 10 (Fig. 4B,C), and plotted against the mean current in the bins. An error-weighted least-squares fit of the linear relationship $\sigma_{\text{exptl}}^2 = \sigma_{\text{filtered}}^2 + \sigma_{\text{baseline}}^2 = \gamma\sigma^2 + \sigma_{\text{baseline}}^2 = \gamma i\mu_I + \sigma_{\text{baseline}}^2$ was computed. (Since baseline points are calculated from 20 ms segments, they receive four times the weight of the 5 ms test bin points). The fitted slopes $\gamma i$ are plotted as a function of membrane potential in Fig. 4D, revealing that the single channel current has an approximately linear current-voltage relationship.

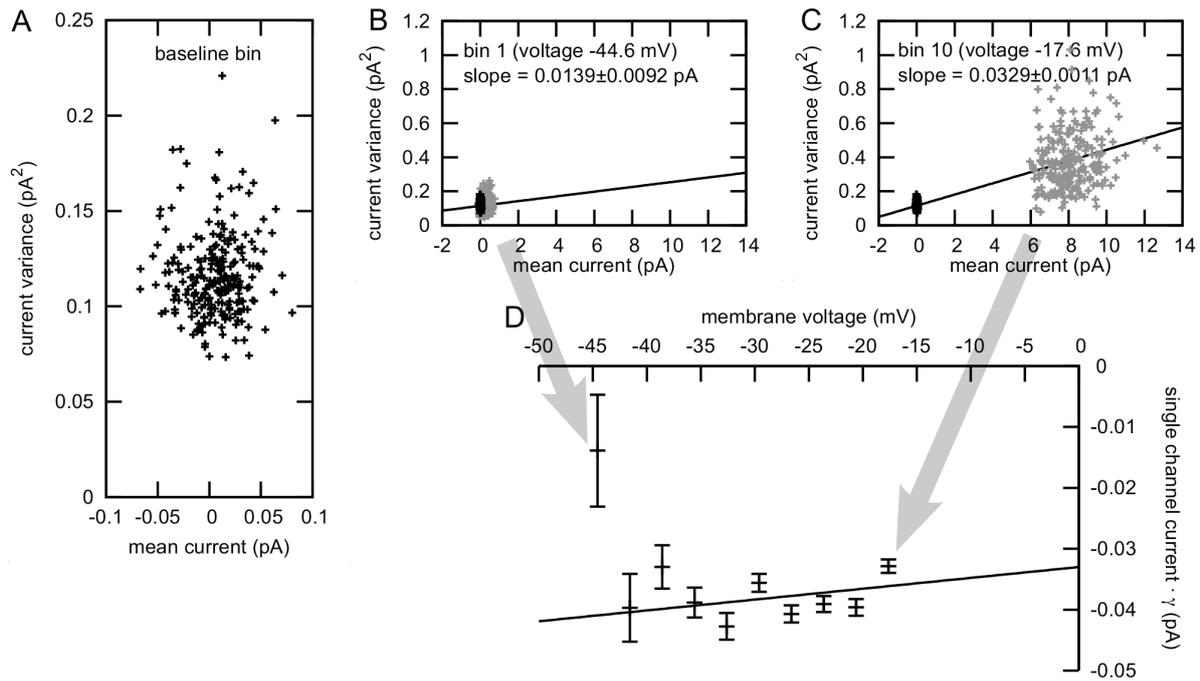

FIGURE 4 Fluctuation measurements in one patch. A) Scatter plot of current variance versus mean current, measured in the baseline bin. B) Same for bin 1 of the ramp (grey points) relative to the baseline (black, same as in A, for comparison). Black line: straight line fit through the baseline and test bin data. The resulting slope estimate and its error are given. C) Same for bin 10. D) Plot of the obtained slope values, with error bars, corresponding to $i \cdot \gamma$, against bin voltage ($i$: single channel current; $\gamma$: correction factor). Straight line indicates error-weighted linear fit, predicting a highly positive reversal potential as expected for calcium channels. - Extracellular $Ca^{2+}$ conc.: 2 mM.

Dividing by $\gamma$ to give the final estimates of single channel current, we plot data for 7 patches at three different calcium concentrations in Fig. 5, as a function of membrane potential. Two patches at 1 mM, two patches at 2 mM, and three at 5 mM were analysed, and straight line fits to the single channel current-voltage relationships were calculated for the pooled data at each concentration. Single channel current is reduced with depolarization, with a shallow slope, implying a highly positive reversal potential, as expected, but the relationship may be better approximated by a Goldman-Hodgkin-Katz current equation than an ohmic linear relationship. However, given the limited voltage range of the data, nonlinear extrapolation is not justified. Table 1 summarizes the parameters of the fits. The single channel current is highly sensitive to the extracellular calcium concentration, as expected (Fig. 5). At 1 and 2 mM, in the physiological range, the single channel currents at −40 mV, just above activation are 0.045 and 0.07 pA respectively. At the peak of the inward current, around −10 mV, the corresponding values are 0.04 pA (1 mM) and 0.065 pA (2 mM). This means that approximately 200 channels are open at the maximum of the ramp response in somatic nucleated patches.

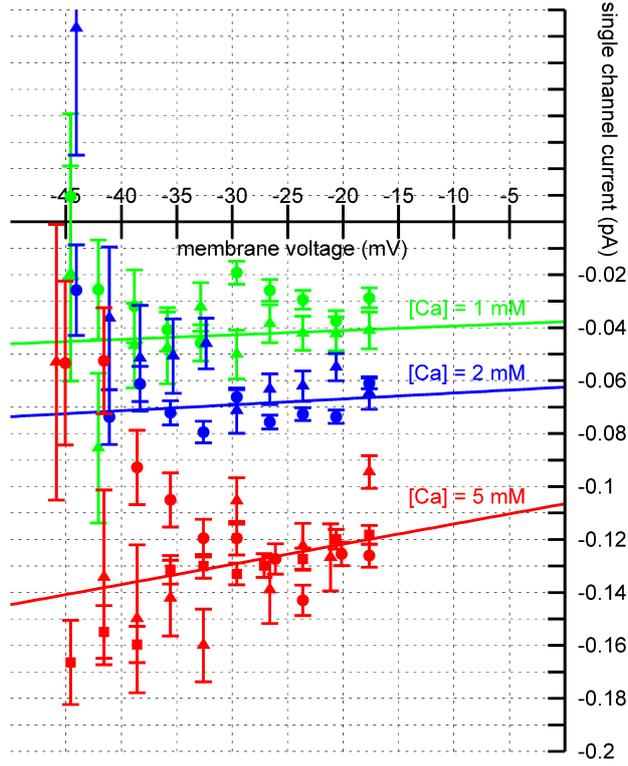

FIGURE 5  Summary of single calcium channel current amplitude, at different membrane potentials and calcium concentrations. Data for 1 mM $[Ca^{2+}]_o$ in green (2 patches), 2 mM in blue (2 patches) and 5 mM in red (3 patches). Data from Fig. 4D are shown as blue filled circles. Different patches indicated by different symbols. Straight line fits of the pooled data at each calcium concentration are shown. There is relatively weak voltage-dependence, but a large increase in single channel current amplitude with increasing extracellular calcium concentration.

TABLE 1  Parameter values for the single $Ca^{2+}$ channel current voltage relationship at different $Ca^{2+}$ concentrations

Single $Ca^{2+}$ channel current-voltage curve (without correction): $\tilde{I} = \tilde{m} \cdot U + \tilde{c}$

Corrected current-voltage curve: $I = m \cdot U + c$

Voltage range: −45 mV to −18 mV

| $[Ca^{2+}]_o$ | $\tilde{m}$ [pA/mV] | $\tilde{c}$ [pA] | $r(\tilde{c},\tilde{m})$ |
|---|---|---|---|
| 1 mM | 0.00009 ± 0.00020 | −0.0203 ± 0.0052 | 0.97 |
| 2 mM | 0.00012 ± 0.00016 | −0.0336 ± 0.0042 | 0.97 |
| 5 mM | 0.00041 ± 0.00019 | −0.0573 ± 0.0050 | 0.97 |
| $[Ca^{2+}]_o$ | $m$ [pA/mV] | $c$ [pA] | $r(c,m)$ |
| 1 mM | 0.00017 ± 0.00037 | −0.038 ± 0.010 | 0.93 |
| 2 mM | 0.00022 ± 0.00030 | −0.0625 ± 0.0089 | 0.83 |
| 5 mM | 0.00076 ± 0.00036 | −0.107 ± 0.012 | 0.67 |

At the extracellular calcium concentrations $[Ca^{2+}]_o$ of 1 mM, 2 mM and 5 mM, the slope $m$ and intercept $c$ of the fitted current ($I$) voltage ($U$) curves (see Fig. 5) are given together with parameter errors and the correlation *r(c,m)* between *c* and *m*. The top set of values ($\tilde{m}$, $\tilde{c}$) correspond to the uncorrected single-channel current, as obtained from straight line fits to data like shown in Fig. 4D. The bottom set of values (*m*, *c*) take into account the correction factor γ (see text).

**Effect of run-down on standard fluctuation analysis**

Here, we provide a quantitative criterion to determine when run-down is so strong that a method like the one suggested above becomes necessary, and when run-down is only mild such that standard fluctuation analysis would work in principle. (However, the problem of low-frequency noise in recordings is a separate one and may favor the approach of high-pass filtering even when run-down is manageable.) Assume *N* recordings of the current $I_j$ (*j*=1...*N*), e.g. *N*=20 to enable sufficiently accurate ensemble averages, and a linear run-down of the number of available channels between successive trials, such that the expectation value of the current $\mu_j$ decreases by $\Delta\mu = \mu_j - \mu_{j+1}$. If one computes the mean of the $I_j$ as if there was no run-down, the expectation value of the result is $\mu$ with $\mu_j = \mu + (\frac{N+1}{2} - j)\Delta\mu$. The variances of the $I_j$ are $\sigma_j^2 = i\mu_j$, where *i* is the single channel current (for the case of small channel open probability). If one now proceeds to estimate the variance from the data by computing $s^2 = \sum_{j=1}^{N} \frac{(I_j - \mu)^2}{N-1}$, again as if there was no run-down, the estimate will be too large, because the residuals ($I_j - \mu$) are taken with respect to the wrong mean. A brief calculation yields the expectation value of $s^2$, $\langle s^2 \rangle \approx i\mu + \frac{N^2}{12}\Delta\mu^2$. The expectation value of the resulting measured single channel current is $\langle i_{\text{measured}} \rangle \approx \frac{\langle s^2 \rangle}{\mu}$. If one wants the systematic relative error due to neglecting run-down to be below 5%, one arrives at the criterion

$$\left|\frac{i_{\text{measured}} - i}{i}\right| = \frac{N^2 \Delta\mu^2}{12 \, i\mu} < 5\% \tag{3}$$

Applied to our dataset, in 2 patches, the channel current stabilized such that only about 10% of the data would have been unusable for standard fluctuation analysis, in 3 patches, about 50% of the data would have been unusable, and in 2 patches, run-down was so strong that standard fluctuation analysis would have been impossible. Note that it is usually the first couple of traces where run-down is fastest, but these are particularly valuable because the channel current is largest, therefore having the best signal to noise ratio, and providing data at large channel numbers.

**Comparison with difference trace method**

Another method for dealing with slow drifts in the experimental conditions has been suggested, based on forming differences of consecutive current traces (35, 36). If traces $I_j(t)$ (*j*=1,2,...) are recorded, one can analyze $I_{1,2}(t) := (I_1(t) - I_2(t))/\sqrt{2}$, $I_{2,3}(t)$ etc. instead of $I_1(t)$, $I_2(t)$ etc. $I_{j,j+1}$ have the same noise spectrum as $I_j$, but any slow drift which is approximately stable between adjacent sweeps will be eliminated. This trick should deal well with normal rundown (unless the number of available channels changes very rapidly or irregularly). However, the method can be affected by low-frequency instabilities that are faster than the time between current traces (like baseline wobble or variations in the leak conductance) just like standard fluctuation analysis.

For comparison, we analyzed the above dataset (seven patches) with the difference trace method. As for the individual traces method, the ramp responses were divided into bins in order to achieve conditions of constant voltage, and the final results were collected in current-voltage plots (Fig. 6A). Here, the data are not high-pass filtered, but only low-pass filtered (Gaussian filter, cut-off frequency $f_c$), and the result depends on $f_c$, since the filter attenuates high frequency components of the channel fluctuation spectrum. We therefore performed the analysis for $f_c$=0.5, 1, 2, 5 and 10 kHz and extrapolated the results to infinite $f_c$, using as a fit function $i_{Ca}(f_c) = i_{Ca}^{\infty} \cdot \frac{2}{\pi}\tan^{-1}(f_c/f_N)$, which is the expected $f_c$-dependence of $i_{Ca}$ for a brick-wall filter ($f_N$: half-power corner frequency of the Lorentzian channel fluctuation spectrum) (35).

The results for the above example patch (cf. Fig. 4) are shown in Fig. 6A. For comparison, the results from the same patch when using the individual traces analysis are shown below (Fig. 6B). At higher membrane voltages, the two methods yield similar results, but at lower voltages, the difference traces method breaks down below about −35 mV, whereas the individual traces methods remains stable up to about −45 mV. The same picture emerged when analyzing the other patches of the dataset. The breakdown points of the two methods were quantified as a deviation of the obtained single channel current from the fit line of 50% or more, and the difference traces method systematically broke down earlier than the individual traces method.

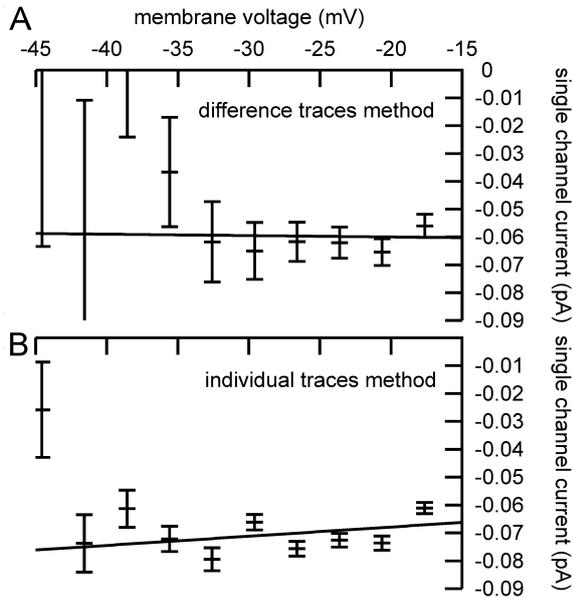

FIGURE 6  Comparison with difference traces method. A) Single calcium channel current at a range of membrane potentials, for one patch, obtained by analysis with the difference traces method. Solid line: straight line fit. B) Results for the same patch when analyzed with the individual traces method (cf. Fig. 4D).

This difference is related to the accuracy with which the slope can be determined in mean-variance scatter plots as shown in Fig. 4B,C. Indeed, in both methods, and for all seven patches, the breakdown points as defined above coincided with the points at which the relative error of the slope estimation from the straight line fit in the scatter plots became 50% or more. (For the difference traces method, the results at $f_c$=1 kHz were investigated.)

The reason for this discrepancy in performance is that the high-pass filtering of the individual traces method improves the signal-to-noise ratio by attenuating the baseline noise while leaving the signal strength relatively unaffected. When estimating the slope in mean-variance scatter plots, the key problem is the scatter of the baseline noise variance, which is also present in the test bins (cf. Fig. 4A,B,C). We determined the scatter of the baseline variance for the two methods, which was reduced by factors around 6 for the individual traces method compared to the difference traces method. E.g. in the patch analyzed above, the baseline noise was 0.20±0.13 pA$^2$ for the difference traces method, but only 0.115±0.020 pA$^2$ for the individual traces method. In contrast, the signal (channel noise variance) is only attenuated by a factor of less than 2 (cf. correction factor $\gamma$). Therefore, especially at lower voltages, where the signal (total current through the $Ca^{2+}$ channels) is smaller, the difference traces method performs worse than the individual traces method.

**Simulated data**

To test the proposed method, we analyzed simulated data. The situation of the example patch above was modeled. A $Ca^{2+}$ channel model $C \rightleftharpoons O \rightleftharpoons D$ (C: closed, O: open, D: deactivated state) was adjusted to fit the experimental current activation time course during the voltage ramp (details: see Supplemental Material), rundown was implemented as a decline of the number of available channels, matching the experimental decay of the $Ca^{2+}$ channel current, and the synthetic channel current traces were generated by simulating the stochastic opening and closing of the channels (Gillespie algorithm, see (18, 37, 38)) and using the experimentally determined single channel current. Care was taken to model the structure of the background noise as accurately as possible. Two noise mechanisms were identified, one which was low-frequency, i.e. long-range correlated at $\tau \approx 300$ ms and was well captured by assuming a fluctuation of the leak conductance $\delta g(t)$, leading to a voltage-dependent current noise $\delta g(t) \cdot U(t)$, and one short-range (details: see Supplemental Material).

The simulated noisy current traces were analyzed with the individual traces method. Fig. 7B shows that the method faithfully reproduces the correct single channel current. For comparison, the simulated data were also analyzed with the difference traces method (Fig. 7A). As for the real data (cf. Fig. 6A,B), both methods yield similar results at higher membrane voltages, but the individual traces method is more accurate (results closer to the real single channel current) and remains stable at lower voltages, where the difference traces method breaks down. Therefore, the positive slope of the single channel current-voltage curve was correctly resolved by the individual traces method (0.00037±0.00015 pA/mV; correct value: 0.00022 pA/mV), but not by the difference traces method (−0.00043±0.00052 pA/mV). As above, we determined the breakdown points of the two methods (−35 mV and < −45 mV), which coincided with the points where the relative error of the slope estimation in the mean-variance scatter plots became 50% or higher. The baseline noise levels were 0.20±0.11 $pA^2$ and 0.110±0.017 $pA^2$, in agreement with the results from the real data.

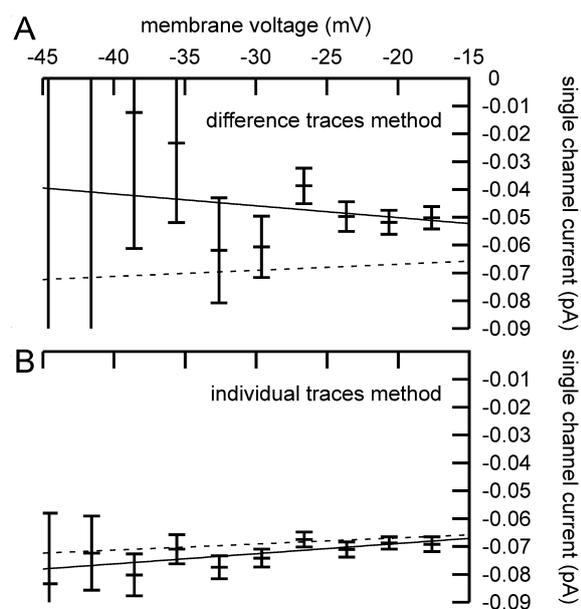

FIGURE 7 Analysis of simulated data. A) Analysis with the difference traces method. Solid line: straight line fit to the data. Dashed line: true single channel current. B) Analysis with the individual traces method.

We also used the simulated data to assess the extent of underestimation of fluctuations resulting from fitting the mean current to the data. The current variance was underestimated by up to 8% (bin 10), but after application of the high-pass filter, the underestimation was reduced to below 0.3%.

## Discussion

In this study, we aimed to measure the single-channel current of native voltage-gated $Ca^{2+}$ channels in neocortical L5 pyramidal neurons, in physiological extracellular $Ca^{2+}$ concentrations. Conventional fluctuation analysis cannot be applied, because of the rapid run-down of $Ca^{2+}$ channel currents. We therefore developed a modified fluctuation analysis protocol which still works under such unstable conditions. Our results provide an essential step towards understanding how many channels participate in physiological calcium currents, how variable these currents are, and how the calcium signaling complex of the channel microenvironment operates. In particular, the results will contribute to a better understanding of stochastic properties of the dendritic $Ca^{2+}$ spike in neocortical L5 pyramidal neurons.

### Measuring native $Ca^{2+}$ channels in neocortical L5 pyramidal neurons

Voltage-gated $Ca^{2+}$ channels consist of one principal subunit $\alpha_1$, which interacts with auxiliary subunits ($\alpha_2\delta$, $\beta$) (7). Channel properties are further modified by accessory proteins and lipid environment and are subject to extensive modulation in a cell-type-specific manner. It is hence important to assess them in native channels *in situ*, rather than in simplified $\alpha_1$ expression systems.

We recorded calcium currents in nucleated patches and encountered current rundown. One approach to counteracting rundown is the perforated patch method ((29) p. 7, p. 45ff), where wash-out of natural intracellular components is slowed. However for nucleated patches, this cannot be applied since a conventional whole-cell configuration needs to be established to be able to aspirate the nucleus (see Methods).

The nucleated patch method obtains somatic membrane (although a small fraction of the membrane may also be derived from the proximal apical dendrite). It should be noted that the method is mechanically more invasive than normal whole-cell measurements, such that some disruption e.g. of the cytoskeleton is possible, which is known to potentially influence channel kinetics (39, 40). Even an alteration of the single channel current cannot be excluded, although we are not aware of studies showing such an effect.

### Measuring single $Ca^{2+}$ channel current amplitude using fluctuation analysis

Accurate estimation of the physiological unitary amplitude of voltage-gated calcium channels has proved a difficult problem. The concentration of the permeant ion is low (1-1.5 mM), while even in near-isotonic (100-150 mM) barium or calcium, single channel conductance is small (7-25 pS, depending on subtype, see (7)). Under the assumption of a linear relationship of single channel conductance to the permeant ion concentration, values in the region of 0.007-0.025 pA might be expected with a driving force of 100 mV and at physiological $Ca^{2+}$ concentration. With mean channel open times in the millisecond or sub-millisecond range, a bandwidth of at least 1 kHz is required to adequately identify channel openings, at which even in optimal patch-clamp recording conditions (41) one has to expect at least 20 fA rms noise. Thus direct resolution of the single physiological calcium channel amplitude is, at best, at the extreme limit of what is achievable with the patch-clamp technique. However, this approach has been attempted, by prolonging channel openings pharmacologically, which allows a more aggressive noise reduction by low-pass filtering (23), or by using low-noise quartz patch pipettes (21, 24). Nevertheless, the accuracy of direct single-channel current measurements remains hard to assess (see discussion below, in *Comparison with previous estimates*).

Another possible approach, as we have adopted here, is to analyze the statistics of current fluctuations in a population containing a large number of channels. Indeed, this was used in an early landmark study on single voltage-gated calcium channels (22). Although fluctuation analysis can in principle measure much lower unitary channel amplitudes, it is important to highlight its limitations. In general, one must assume that all channel openings are identical in amplitude – if not, the single channel current estimate will be an average for the different subtypes of channel openings, weighted according to their prevalence, although this may still be useful information. Fluctuation analysis also assumes independence of channels in the population – however this is usually a valid assumption. In conventional fluctuation analysis, the expectation current is estimated as the ensemble mean current, by averaging many identical trials. However, this approach has the major drawback of requiring a long period of stationarity, to accumulate at least 20 to 30 responses under identical conditions. We found that this was not realizable in this preparation, and may be a questionable strategy to apply to calcium channels under any circumstances, owing to the progressive rundown to which they are prone, even when ATP and GTP are provided in the pipette solution (26, 27). Therefore, we adapted this method in several ways. First, we took the approach of fitting individual responses to a smooth function which was sufficiently flexible to follow the form of macroscopic currents accurately without over-fitting, i.e. following the time course of fluctuations. This resembles previous fitting methods applied to synaptic currents (33, 34). In both cases, fitting a function by least-squares can result in a slight underestimation of the variance – i.e. the fitted function is closer overall, in a least-squares sense, to the response than is the underlying expectation current. However, this kind of error is reduced by using a reasonably extended trial, essentially allowing time for a high number of fluctuations around the mean within each trial. In the present situation, we found in simulations that the underestimation was up to 8%. Applying a high-pass filter to the fluctuation trace dealt with this problem in a controlled way, reducing the error to below 0.3%. Secondly, the high-pass filter rejected extraneous low-frequency artifacts, which we found to be a useful enhancement to the robustness of the method. We showed how to compensate for this step, taking into account the temporal correlation of the fluctuations (Fig. 3A). We verified the proposed method by simulation, showing that it faithfully recovers the single-channel current.

Finally, in the present experimental situation, ramp stimuli had a number of advantages over voltage steps. Firstly, leak subtraction is easily done in individual sweeps by fitting a straight line to the initial part of the ramp, such that the sampled baseline and leak are maximally close to the measurement range. Also, there is no capacitative transient to worry about, only a step capacitative current, most of which is cancelled by the amplifier circuitry, and any residual current is eliminated by the leak subtraction. For voltage steps, one would have to work with additional steps to hyperpolarized and/or slightly depolarized voltages and subtract the linearly scaled response from the test traces to correct for leak and residual uncancelled capacitative currents. Secondly, a ramp stimulus leads to a smooth, gradual time course of the mean current response, which can be well fitted with a heuristic function. Step responses are more peaked and therefore more prone to overfitting. The overfitting-correction with a high-pass filter only works well if the mean current contains only low frequencies (cf. Results section *Fluctuation analysis of voltage ramp responses*), which is not the case for a brief peak. Thirdly, with ramps, the full voltage range where channels are activated is automatically and efficiently sampled with one generic stimulus, whereas for steps, one would have to pick specific test voltages to resolve the voltage dependence of the single-channel current. However, in other situations, particularly for fast inactivating channels, and when a wide range of sampled $p_{open}$ values is desired (e.g. $Na^+$ channels), step stimuli could be a better choice, and the presented method can also be applied there, although the results may be less reliable in places where the mean current changes rapidly.

The method we developed here for calcium channels should be useful in the study of other ion channel types, particularly in situations where channels run down quickly and baseline and linear leak are not perfectly stable. We provided a quantitative criterion when run-down is too strong so that standard fluctuation analysis becomes systematically flawed. Moderate run-down can also be eliminated by analyzing the differences of consecutive traces, as has been suggested before (35, 36). However, the high-pass filtering of the presented method has the additional advantage of attenuating other low-frequency noise. In our dataset, this gave rise to more accurate results, and better stability of the method at lower signal strength (low channel current at voltages where only few channels are activated), compared to the difference traces method.

We assumed that channel open probability $p$ is small, such that the linear approximation of Eq. 2 is valid. The maximal error from this approximation is given by $p$, i.e. if $p$ is 10%, one underestimates the single channel current by up to 10%. (If data at smaller $p$ are also available, the error decreases.) For $Ca^{2+}$ channels, at the voltages of this study (below $-18$ mV), the channel open probability has been measured to be small, below 0.2 even at the top end of the voltage range (25), such that an error above 10% arising from this source is unlikely. If an ion channel type has larger open probabilities (as would for example be the case for Na channels, see (4, 5)), the method presented here is not valid, because the number of channels $N$, which decreases due to run-down, does not disappear from the mean-variance equation. However, it could be extended to the use of the full quadratic relationship (Eq. 1) for channel types for which this proves necessary, or experimental conditions could be adjusted such that one is again in the low $p_{open}$ regime.

## $Ca^{2+}$ channel subtypes in neocortical L5 pyramidal neurons

Of which subtype are the $Ca^{2+}$ channels encountered in neocortical L5 pyramidal neurons? Eleven subtypes of the principal channel-forming $\alpha_1$ subunit are known, distinguished by electrophysiological and pharmacological properties and grouped in three gene families (7). The first family $Ca_V1$, encodes L-type channels, while $Ca_V2$ genes encode N-type, splice variants P- and Q-type, and R-type channels, and the $Ca_V3$ genes encode T-type channels. $Ca_V1$ and $Ca_V2$ channels require higher depolarisations to activate than $Ca_V3$ channels and are therefore called high-voltage activated (HVA) calcium channels, while $Ca_V3$ channels are called low-voltage activated (LVA).

In (27), the pharmacology of calcium currents in nucleated patches from neocortical L5 pyramidal neurons was examined and no T-type channels were found, but all other subtypes, L-, N-, R-, P- and Q-types were inferred from pharmacology. In (42), $Ca^{2+}$ channels in dendrites and somata of hippocampal pyramidal neurons were studied. In dendrites, this study found only occasional L-type channels, but predominantly a class of HVA medium conductance channels which was tentatively identified with R-type, and LVA T-type channels were frequently encountered. In somata, a larger density of L-type channels and fewer R-type channels were found.

Our own pharmacological data (see Supporting Material), while preliminary, suggest that the bulk of the $Ca^{2+}$ channel current we measured in somatic nucleated patches was due to R-type and/or Q-type channels, with a small contribution from L-type channels. The activation threshold of our currents was much more depolarized than typical for LVA T-type channels. Moreover, the single exponential component of the autocorrelation of fluctuations (Fig. 3A) suggests a reasonably homogeneous population of channels being activated under our conditions. (For more detailed discussion, see Supporting Material.). However, in general, the values that we have measured reflect the overall average current amplitude of the subtypes present in the cell which are activated by the particular voltage stimulus used, weighted in proportion to their contribution to the total current.

**Comparison with previous estimates of single Ca$^{2+}$ channel current**

The calcium concentration in the cerebrospinal fluid of rats is 1.0 to 1.5 mM ((17), see also note in (21)), while a standard value for artificial cerebrospinal fluid recipes that is often used in brain slice experiments (e.g. (11, 30)) is 2 mM. As for the single channel current in extracellular barium $i_{Ba}$, $i_{Ca}$ depends on the calcium channel subtype, but there is no uniform ratio between $i_{Ba}$ and $i_{Ca}$. T-type channels produce about the same current in barium and in calcium, whereas L-type and N-type channel currents are larger in barium (43). Hence the recognized hierarchy of single channel currents in barium, Ca$_V$1 (L-type) > Ca$_V$2 (N,P,Q,R-type) > Ca$_V$3 (T-type), is not valid in calcium. In (21), $i_{Ca}$ for L-, N- and T-type channels was measured and the current hierarchy N-type > L-type > T-type was obtained, with the N-type current 38% larger than the L-type current, and the T-type current 16% smaller than the L-type current.

$i_{Ca}$ does not depend linearly on the extracellular calcium concentration [Ca$^{2+}$]$_o$ or on the membrane voltage. $i_{Ca}$ saturates at high [Ca$^{2+}$]$_o$ and the dependence can be described with the Hill equation (see discussion in (21)). The voltage-dependence of $i_{Ca}$ follows the nonlinear Goldman-Hodgkin-Katz current equation ((6) p. 445ff). Therefore, we restrict our comparison to $i_{Ca}$ values at the same extracellular calcium concentration and membrane voltages, rather than comparing slope conductances. In (22), calcium channels in bovine adrenal chromaffin cells were studied. In these cells, half of the calcium current is thought to flow through P/Q-type channels, and the other half through N- and also L-type channels (reviewed in (36)). At a membrane potential of −12 mV, fluctuation analysis yielded a single channel current of 0.03 pA in 1 mM Ca$^{2+}$ and 0.09 pA in 5 mM Ca$^{2+}$, which is in good agreement with our results (0.04 pA in 1 mM Ca$^{2+}$, 0.11 pA in 5 mM). In (23), L-type calcium channels were studied in arterial smooth muscle and channel openings were prolonged with BayK8644, which resulted in channel open times of 10 ms and longer, allowing filtering at 500 Hz. In cell-attached recordings in 2 mM [Ca$^{2+}$], 0.10 pA at −20 mV and 0.20 pA at −40 mV were obtained. Compared with our almost constant value of 0.07 pA in the range between −20 and −40 mV, however, this current-voltage relationship is much steeper. In fact, extrapolating the values beyond −20 mV predicts a reversal potential of about 0 mV, although E$_{Ca}$ should be highly positive. Taking into account that our currents are probably predominantly due to Ca$_V$2 channels and should hence be about 40% larger than L-type currents under otherwise equal conditions, the values in (23) seem comparatively large. In (21), a direct single-channel recording approach was also used, with quartz pipettes employed to reduce noise, and like in (23), larger values were obtained than the ones reported here. For N-type channels in 2 mM Ca$^{2+}$, 0.21 pA at −25 mV was reported. Compared with our value of 0.068 pA, this is about a factor of 3 larger. Part of the discrepancy may be due to different channel subtypes present in layer 5 pyramidal neurons. For example if R-type channels predominate, they might have a smaller $i_{Ca}$ than N-type channels, despite both being members of the Ca$_V$2 family. L-type channels may also contribute, which in (21) are reported to have a slightly smaller $i_{Ca}$.

Thus, single channel current estimates obtained by direct single-channel recordings appear to yield larger values than fluctuation analysis. When trying to judge by eye the size of a single channel current step at a low signal-to-noise ratio, close to the resolution limit, there is a danger of overestimating the single-channel current, if only a small subset of events are large enough to detect, or if insignificant fluctuations are mistaken for channel openings (see discussion in (19)). In (22), single-channel currents in 95 mM Ba obtained by single-channel recordings were compared with fluctuation analysis results, yielding about 40% larger values in direct single-channel recordings. As a possible reason, a loss of noise power due to filtering was discussed, which would reduce the single-channel current estimate from fluctuation analysis. However, we fully corrected for the effects of filtering, so this should not be an issue. Also discussed was the possibility of heterogeneity in the channel population, i.e. that small channel openings can be missed by direct single-channel recordings but are picked up by fluctuation analysis.

The relaxation time constants $\tau$ we obtained from autocorrelations were in the range 0.2-0.8 ms. This is consistent with mean channel open times between 0.4 and 1 ms reported in (42), obtained by single-channel recordings in high extracellular barium.

**Functional implications**

In computational models of ion channels, it is usual to consider the limit of many ion channels, leading to deterministic differential equations for neuronal dynamics (45). Increasingly, though, it is appreciated that the stochastic current fluctuations due to the opening and closing of individual ion channels are not negligible, for example they can lead to spontaneous action potentials and have a large effect on action potential timing (18). In (20), a detailed computational study of stochastic calcium spikes in cerebellar Purkinje cell dendrites was performed, and it was found that large variability in calcium spike bursts, also seen experimentally, is produced, in a model, by the interaction of stochastic calcium influx with downstream calcium-dependent conductances, via stochastic intracellular calcium transport and buffering. A single channel permeability of 2.5 x $10^{-5}$ $\mu m^3$ $ms^{-1}$ was assumed for P-type channels, resulting in a single channel current of about 0.035 pA (at body temperature, 2 mM extracellular calcium and −50 mV membrane voltage), which is close to our values for 1 mM extracellular calcium.

The extreme sensitivity of calcium nanodomain concentrations to the calcium influx (46) means that downstream biochemical signaling is also critically dependent on the exact value of $i_{Ca}$. For example, stochastic calcium influx is predicted to have a major effect on CaMKII activation in dendritic spines (47) and therefore potentially on long-term potentiation. Although some of this calcium enters through NMDA receptors, dendritic calcium channels will also contribute, and therefore their physiological unitary properties, as well as those of NMDA receptors, are relevant in this regard.

To our knowledge, there have been no studies which directly recorded single voltage-gated calcium channels in physiological calcium concentration in the apical dendrite of layer 5 pyramidal neurons. Based on current-clamp and ionic substitution and blocker experiments, it has been suggested that there may be a "hot spot" of $Ca^{2+}$ channel density at the dendritic $Ca^{2+}$ spike initiation zone (12, 48). It is conceivable that the subtype composition of dendritic calcium channels is different to that in the soma, but in the absence of further evidence, our data give the best available indication of the physiological single-channel current amplitude in the dendrite as well as the soma.

# Conclusion

By adapting and refining a fluctuation analysis approach, we have been able to provide an improved estimate of the single channel current $i_{Ca}$ of $Ca^{2+}$ channels in physiological levels of extracellular $Ca^{2+}$, a quantity which has rarely been measured, despite its importance for stochastic single-channel effects or calcium nanodomains. We further measured the timescale of autocorrelation of calcium channel gating. These results should assist in building more realistic computational models to understand calcium-dependent signaling in neurons.

## Author contributions
C.S. and H.P.C.R. designed the experiments. C.S. performed the experiments and analyzed the data. C.S. and H.P.C.R. performed the simulations and wrote the manuscript.

## Acknowledgments
C.S. acknowledges funding by the German Academic Exchange Service (DAAD), the Biotechnology and Biological Sciences Research Council (BBSRC), the Cambridge European Trust (CET) and the Research Innovation Fund of the University of Freiburg.

## Supporting citations
References (49, 50) appear in the Supporting Material.

# Supporting material to:
# "Fluctuation analysis in nonstationary conditions: single Ca channel current in cortical pyramidal neurons"

This supplementary document contains more detailed information on the fitting function to voltage-ramp evoked calcium currents, the filter correction, pharmacological results on the observed $Ca^{2+}$ channel currents and details on the modeling of $Ca^{2+}$ channels and the background noise.

## Fitting of current responses to voltage ramp stimuli

Current responses to voltage ramp stimuli were fitted with a piecewise polynomial, consisting of a flat segment and two linear segments linked by cubic splines (Fig. S1), giving a smooth heuristic function whose eight parameters were fitted by minimizing squared deviation from the data, which was taken as the estimate of the mean current $\mu_I$ for fluctuation analysis.

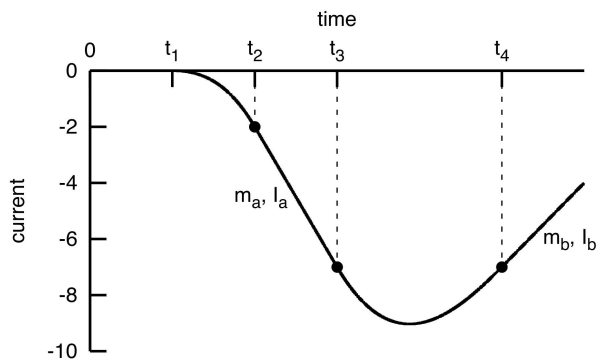

FIGURE S1  Fit function for voltage-ramp-evoked calcium currents. A heuristic piecewise polynomial function with continuous derivative was fitted by least squares to individual leak-subtracted ramp responses (see main text, Fig. 2A). The function consisted of an initial flat segment and two linear sections joined by cubic splines. The transition times $t_{1...4}$, gradients $m_{a,b}$ and intercepts $I_{a,b}$ were all free parameters.

# Correction for effects of filtering

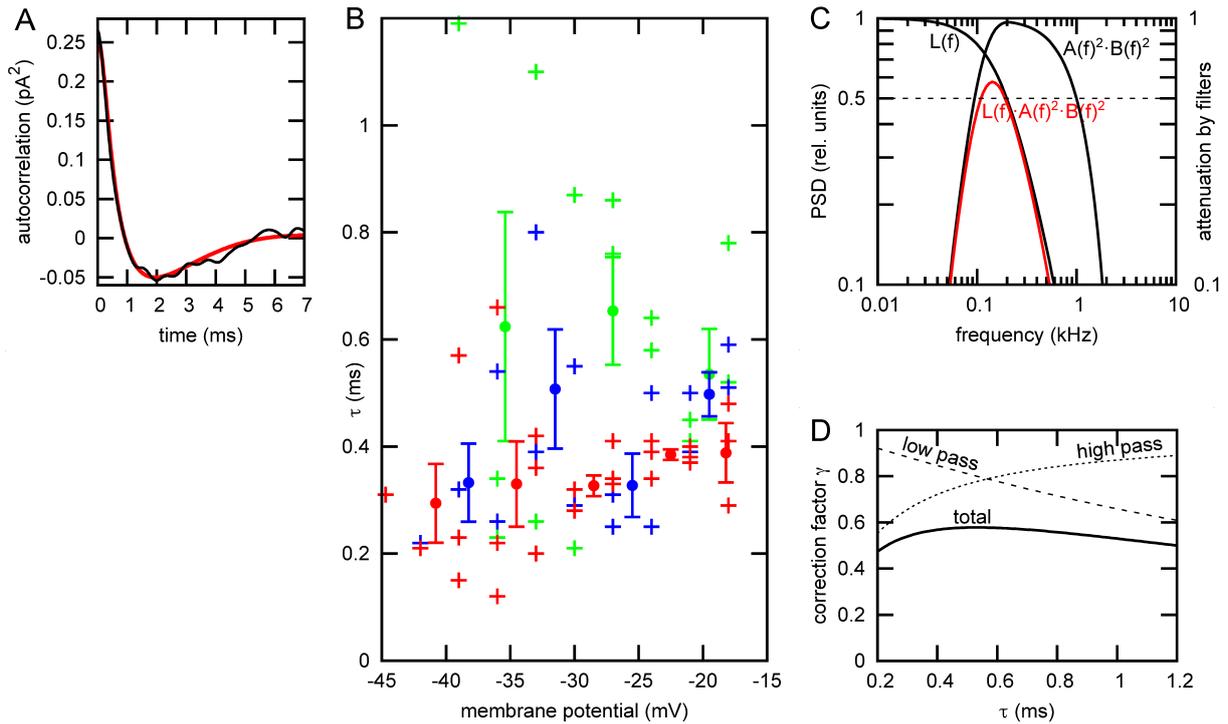

FIGURE S2  Correction for effects of filtering. A) Channel noise is dominated by a single exponential relaxation. Black trace shows the autocorrelation of the signal computed in one time bin (bin 10), with baseline autocorrelation subtracted. The red curve shows the least-squares fit to a model of exponentially-relaxing fluctuations (single Lorentzian spectral component). Undershoot is caused by the high-pass filter. B) Mean channel open/burst time $\tau$, the relaxation time constant of fluctuations, extracted by this procedure is plotted for multiple patches and time bins, as a function of membrane potential. Green is 1 mM [$Ca^{2+}$], blue is 2 mM and red is 5 mM. Filled circles with error bars are mean and standard error of the mean obtained from about five adjacent data points. $\tau$ values are concentrated in the range 0.2-0.8 ms. C) Illustration of the effect of the bandpass filtering in the frequency domain. $L(f)$ indicates the Lorentzian distributed power of the underlying current fluctuations (exponential noise of $\tau = 0.8$ ms, therefore Lorentzian half-power corner frequency = 200 Hz; PSD: power spectral density). The combined effect on noise power of the low- and high-pass filter characteristics ($A(f)$ and $B(f)$ respectively) is shown as the curve $A(f)^2 \cdot B(f)^2$. Channel noise power which passes through filtering is shown in red: $L(f) \cdot A(f)^2 \cdot B(f)^2$. D) The ratio of the integral of the noise power density passed by the filter to the total noise power gives the correction factor $\gamma$, whose dependence on the noise relaxation time constant $\tau$ is shown as the solid curve, the combined effect of the low-pass filtering (dashed curve) and high-pass filtering (dotted curve). $\gamma$ is quite insensitive to $\tau$ over the experimentally-relevant range.

## Methods

The autocorrelation $R(t)$ of a current trace $I(t)$ for a time bin (Fig. S2A) was computed as the mean of $I(t_1) \cdot I(t_2)$ with $t_2 - t_1 = t$ and at least one of $t_1$ and $t_2$ inside the bin. Hence the autocorrelation could also be estimated for $t$ larger than the bin width of 5 ms.

# Pharmacology of observed Ca$^{2+}$ channel currents

## Methods

### Solutions, drugs, chemicals

Nifedipine (L-type Ca$^{2+}$ channel blocker) was obtained from Sigma (Gillingham, U.K.). A 40 mM stock solution in ethanol was produced on the day of the experiment. (S)-(−)-Bay K 8644 (L-type activator) was obtained from Tocris (Bristol, U.K.). A 10 mM stock solution in ethanol was prepared, kept in the fridge and used within two days. ω-Conotoxin GVIA (N-type blocker) and ω-agatoxin IVA (P-type blocker) were obtained from Tocris. 100 μl of each drug dissolved in HEPES-buffered extracellular solution was applied with a pipette into the recording chamber, and the bath solution was gently stirred by aspiration with the pipette several times. Since the recording chamber solution volume was around 3 ml, the concentration of the drug was diluted by a factor of about 30.

### Recording

Ca$^{2+}$ channel currents were recorded during depolarizing ramps, e.g. from a holding potential of −70 mV rising to 0 mV (slope 0.6 mV/ms, stimulus repeated every 3 s), and also during depolarizing voltage steps (e.g. holding potential −70 mV, 100 ms long steps to test potentials cycling through −80 mV, −60 mV, −30 mV, −20 mV, −10 mV and 0 mV, stimulus repeated every 3 s). In nifedipine block experiments: in 6 of 8 patches, ramp stimulus with a holding potential of −70 mV, prepulse at −90 mV for 200 ms, rise to 0 mV, stimulus repeated every 5 s. In one patch, holding potential of −60 mV, no prepulse, depolarization to 0 mV, repeated every 2 s. In one patch, holding potential of −70 mV, no prepulse, depolarization to 0 mV, repeated every 2 s.

### Data analysis

Leak subtraction for step responses (Fig. S2 E) was carried out by subtraction of scaled ensemble-averaged responses to small steps (holding potential +/− 10 mV) below the activation range of the calcium current.

## Results: Pharmacology of Ca$^{2+}$ channels in nucleated patches from neocortical L5 pyramidal neurons

Ca$^{2+}$ channel currents were recorded during depolarizing ramps, and also during depolarizing voltage steps. Application of 200 μM cadmium completely blocked the current (see main text, Fig. 1C; $n = 11$ cells). The block could not be reversed by perfusing the bath again with Cd-free solution (n=2). Partial block was observed at 20 μM Cd$^{2+}$ (n=2), which was reversible (n=1). 2 μM Cd$^{2+}$ did not block the current (n=1).
In most cases (5 of 8 patches), nifedipine (L-type Ca$^{2+}$ channel blocker) produced no clear blocking effect (Fig. S2 A), but a partial block was noted in 3 patches (Fig. S2 D). Generally, a small partial block was difficult to differentiate from the natural time course of rundown. Therefore, the Ca$^{2+}$ channel current was plotted against time (Fig. S1 A-D), and if addition of the blocker resulted in a drop of the current below what would be expected from the extrapolated rundown beforehand, the data were interpreted as showing a block. Nifedipine was used at several concentrations: 3 μM (2 patches, partial block in 1), 30 μM (4 patches, partial block in 1), 50 μM (1 patch, partial block) and 70 μM (1 patch, no block). Bay K8644 (1 μM), an L-type channel agonist, had no discernible effect on the ramp-activated inward current (n=2), but boosted the current during depolarizing voltage steps (n=2, Fig. S2 E). We conclude that some L-type channels are present in somata of L5 pyramidal neurons, but they do not account for the major component of the calcium current. ω-Conotoxin GVIA (N-type

Ca$^{2+}$ channel blocker) at 1 µM generally did not block the Ca$^{2+}$ channel current (Fig. S2 B, n=7, only in one patch possible slight block). ω-Agatoxin IVA (P-type Ca$^{2+}$ channel blocker, 30 nM) also had no blocking effect (Fig. S2 C, n=4).

In conclusion, the full block by Cd$^{2+}$ shows that the inward current observed in ramps is carried by voltage-gated calcium channels. The lack of block by ω-conotoxin GVIA and ω-agatoxin IVA indicates that N and P type channels are not involved, while T-type channels would have a much lower activation threshold (1), suggesting that the ramp-evoked current corresponds mostly to R-type (Ca$_V$2.3), or possibly Q-type (Ca$_V$2.1) channels, with a small contribution from L-type channels.

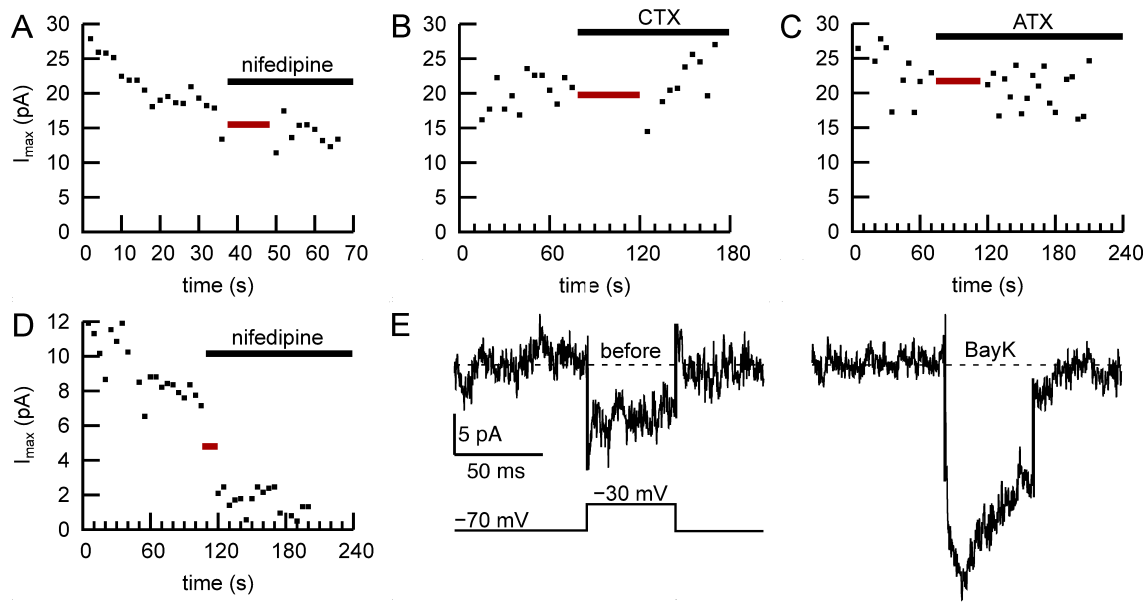

FIGURE S2  Pharmacology of Ca$^{2+}$ channels. In the top row, three example patches are shown in which peak current was insensitive to A) nifedipine (30 µM), B) ω-conotoxin GVIA (CTX, 1 µM) and C) ω-agatoxin IVA (ATX, 30 nM). D) Patch in which Ca$^{2+}$ channel current was partially blocked by nifedipine (3 µM). Red bars indicate wash-in period during which currents were not measured, black bars indicate the presence of blocker. Note the quite typical time course of rundown of the calcium channel current in panel A. E) Patch in which BayK (1 µM) boosted the current during depolarizing voltage steps.

## Discussion: Ca$^{2+}$ channel subtypes in neocortical L5 pyramidal neurons

We concluded that the current which we studied in voltage ramp responses in nucleated patches was likely to be due to high-voltage-activated R- and/or Q-type channels, with a small contribution from L-type channels. Cadmium (200 µM) totally blocked the current. Nifedipine (3-70 µM), an L-type blocker, only resulted in a partial block in some patches, and Bay K8644 (1 µM), an L-type agonist, only boosted the current in some patches. The toxin blockers ω-conotoxin GVIA (1 µM), which fully blocks N-type channels (2), or ω-agatoxin IVA (30 nM), which is reported to fully block P-type channels, did not affect the current. We did not test concentrations of ω-agatoxin IVA which would be required to block Q-type channels (K$_d$ = 100-200 nM, see (3)). The activation threshold of the current was much more depolarized than typical for T-type channels. Our results differ somewhat from those of a previous study (4), in which the pharmacology of calcium currents in nucleated patches from neocortical L5 pyramidal neurons was also examined, but using barium as the permeant ion. There, also no T-type channels were found, but the presence of all other subtypes, L-, N-, R-, P- and Q-types was reported: nifedipine, ω-conotoxin GVIA, ω-agatoxin IVA, as well as ω-conotoxin MVIIC (blocks Q-, N- and P-type) and SNX-482 (may block R-type) all blocked fractions of the current. In (5), $Ca^{2+}$ channels were studied in dendrites and somata of hippocampal pyramidal neurons, again with barium as the permeant ion. In dendrites, consistent with the results reported here, only occasional L-type channels were found, but predominantly a class of HVA medium conductance channels which was tentatively identified with R-type, based on the resistance of the channels to ω-conotoxin MVIIC. However, in somata, ω-conotoxin MVIIC was found to have a significant blocking effect, and LVA T-type channels were frequently encountered. The reasons for these discrepancies are not clear, and may well partly be due to our use of physiological calcium as the permeant ion, the different cell type studied in (5) and the difficulty of performing clean $Ca^{2+}$ channel pharmacology due to rapid $Ca^{2+}$ channel rundown and the lack of highly subtype-specific $Ca^{2+}$ channel blockers.

## Modeling of $Ca^{2+}$ channels and background noise

### $Ca^{2+}$ channel model

We used a model with one closed, one open and one deactivated state:

$$C \underset{k_c(U)}{\overset{k_o(U)}{\rightleftharpoons}} O \underset{k_d^-}{\overset{k_d^+}{\rightleftharpoons}} D \qquad (1)$$

$k_d^+$ and $k_d^-$ were chosen such that the deactivation time constant is 70 ms, and the current deactivates to 10% of the peak value (cf. Fig 1A of main text):
  $k_d^+$=0.015 kHz
  $k_d^-$=0.0015 kHz

For $k_c(U)$, an exponential voltage dependence was used:
  $k_c(U) = B \cdot \exp(-\beta \cdot U)$

The parameters were chosen such that the mean channel open time is 0.4 ms at −30 mV and 0.8 ms at 0 mV (cf. Fig. 3B of main text):
  $B$=1.25 ms$^{-1}$
  $\beta$=0.023 mV$^{-1}$

$k_o(U)$ was chosen such that the experimental current time course in response to the ramp was reproduced. To achieve this, a sigmoidal voltage dependence was necessary:

$$k_o(U) = \frac{A}{1+K \cdot \exp(-\alpha U)}$$

Furthermore, an open probability of 1% was imposed at the maximum of channel activation within the analyzed data range (90 ms after onset of the ramp). This value is arbitrary, but was chosen such that the linear approximation of the relationship between current mean and variance in fluctuation analysis is valid (see main text). The parameters were obtained by a least squares fit to the experimental current time course:
  $A$=0.0215 kHz
  $K$=0.00983
  $\alpha$=0.167 mV$^{-1}$

### Background noise

To characterize the background noise present in our recordings, we plotted autocovariances $R(t,t+\Delta t)=\langle I(t) \cdot I(t+\Delta t)\rangle$ of the data, centered at different times $t$. On longer timescales, the autocovariances decayed exponentially with a time constant of about 300 ms, but in the course of the voltage ramp, the autocovariance values became progressively smaller than expected. This can be explained by assuming a leak current $I_{leak}(t)=g(t)\cdot U(t)$ which reverses at 0 mV and has a fluctuating conductance $g(t)=g_0+\delta g(t)$. The noisy $\delta g(t)$ traces with exponential autocovariance were generated by convolving white Gaussian noise with an exponential function.

To capture the remaining short-range (i.e. high frequency) noise of the data, the power spectral density PSD($f$) ($f$: frequency) was obtained from the baseline bin (width: 20 ms). Then, the expected $PSD_{l-r}$ for the above long-range noise was computed (which was concentrated almost exclusively in the zero-frequency component), and subtracted from the experimental PSD, to obtain $PSD_{s-r}$ of the remaining short-range noise. The noise traces were then generated by convolving white Gaussian noise with the Fourier transform of $\sqrt{PSD_{s-r}(f)}$.

## Supporting references